# Computational complexity of three-dimensional Ising spin glass: Lessons from D-wave annealer

Hao Zhang[1,*] and Alex Kamenev[1,2]
[1]*School of Physics and Astronomy, University of Minnesota, Minneapolis, Minnesota 55455, USA*
[2]*William I. Fine Theoretical Physics Institute, University of Minnesota, Minneapolis, Minnesota 55455, USA*

Finding an exact ground state of a three-dimensional (3D) Ising spin glass is proven to be an NP-hard problem (i.e., at least as hard as any problem in the nondeterministic polynomial-time (NP) class). Given validity of the exponential time hypothesis, its computational complexity was proven to be no less than $2^{N^{2/3}}$, where $N$ is the total number of spins. Here, we report results of extensive experimentation with D-Wave 3D annealer with $N \leqslant 5627$. We found exact ground states (in a probabilistic sense) for typical realizations of 3D spin glasses with the efficiency, which scales as $2^{N/\beta}$ with $\beta \approx 10^3$. Based on statistical analysis of low-energy states, we argue that with an improvement of annealing protocols and device noise reduction, $\beta$ can be increased even further. This suggests that, for $N < \beta^3$, annealing devices provide most efficient way to find an exact ground state.

## I. INTRODUCTION

Optimization problems are ubiquitous across science, technology, and industry [1,2]. A large class of such problems can be formulated as a task of finding a bit string, $\{\sigma^z\} = \{\pm 1, \pm 1, \ldots, \pm 1\}$, of length $N$, which minimizes a certain cost function, $H[\{\sigma^z\}]$. While the latter can, in principle, assume an arbitrarily complicated form, many studies [3–13] restrict it to a quadratic form, which mimics binary interactions of Ising spins:

$$H[\{\sigma^z\}] = \sum_{i<j} J_{ij} \sigma_i^z \sigma_j^z. \tag{1}$$

Here, an $N \times N$ matrix $J_{ij}$ encodes coupling strengths between the spins. A physically and conceptually important example is provided by the Edwards-Anderson (EA) model of a spin glass [8,14–18], where $J_{ij}$'s are restricted to a lattice in $D$ spatial dimensions, and are randomly and independently drawn from a distribution with zero mean and width $J$. Hereafter, we put $J = 1$ and thus measure the energy (i.e., the cost function) in this dimensionless unit.

It was proven [19–21] that finding a spin configuration, out of $2^N$ possibilities, exactly minimizing the EA energy in $D > 2$ is an NP-hard problem, meaning it is at least as hard as any problem in the nondeterministic polynomial-time (NP) class [22,23]. This means that no known algorithm (classical or quantum) [18,24–54] can find or verify an answer in a polynomial time. For $D = 3$, it was proven [55] that no algorithm can be more efficient than $2^{N^{2/3}}$ in the $N \to \infty$ limit, provided *exponential time hypothesis* (ETH) [56] holds. We provide a simple illustration of $N^{2/3}$ algorithm in Sec. V. These results do not limit the possibility of an algorithm (or an analog device) with the exponential scaling, $\sim 2^{N/\beta}$, which is more efficient for $N < \beta^3$. Here, $\beta$ is a constant specific to a given hardware and software implementation of a computation procedure [57,58]. The goal of this paper is to discuss whether there are fundamental physical limits on $\beta$, based on extensive experimentation with the D-Wave three-dimensional (3D) annealer [47–54].

Despite its 50-year history, the physics of the 3D EA model is not yet fully understood. The debate [59–70] is between the replica symmetry breaking scenario, the droplet picture, and the intermediate scenario that combines between the two [62,71,72]. A definitive delineation between them is rather tricky and may require extremely large system sizes and extensive statistics. This is not the goal of this communication. Instead, we present a detailed statistical analysis of low-energy states, reachable with the help of annealing protocols. We argue that having a sufficiently large set of such states and employing a postprocessing classical algorithm, one may find the true unique (up to global spin reversal) ground state—i.e., the absolute best optimization outcome.

Analysis of the computational complexity of this procedure is based on counting the number of local minima (or rather *basins* of attraction, defined below) with the excess energy $\delta = E - E_0$, separated by a Hamming distance of order $N$ from each other. According to our data, this number scales as

$$m(\delta, N) \propto \exp\left(\frac{\delta}{2\delta_0}\right). \tag{2}$$

Here, $E$ is an energy of a deep minimum, $E_0$ is the ground-state energy, and $\delta_0 = -E_0/N > 0$ is the ground-state energy per spin (for the box distribution, $-1 < J_{ij} < 1$, and the D-Wave Advantage architecture, $\delta_0 \approx 1.6$). Using the D-Wave annealer and the cyclic annealing protocol [73,74], explained below, we generate a large ensemble of low-energy states. Their average excess energy $\delta$ appears to scale linearly with the system size $N$,

$$\delta \approx N/\beta_{\text{eff}}, \quad (3)$$

where $1/\beta_{\text{eff}}$, the average excess energy per spin, may be loosely identified with an effective temperature of the generated ensemble. We then show that, given sufficiently many states per basin, one can digitally "cool" the ensemble down to the true ground state (with a probability, which may be consistently increased by increasing the size of the ensemble of states). The computational effort involved in this procedure scales as $\sim m(\delta, N) \propto e^{N/(2\delta_0 \beta_{\text{eff}})}$, focusing on the exponent. This leads to $\beta = (2\delta_0 \log 2)\beta_{\text{eff}} \approx 2.2 \beta_{\text{eff}}$. Therefore, the computational complexity is tight to the inverse effective temperature $\beta_{\text{eff}}$ of an available sufficiently large ensemble of low-energy states. The D-Wave annealer and cyclic annealing protocol allow us to reach $\beta_{\text{eff}}$ in excess of $10^3$ with 10 ms time per one low-energy state.

*Contact author: hao.zhang.quantum@gmail.com

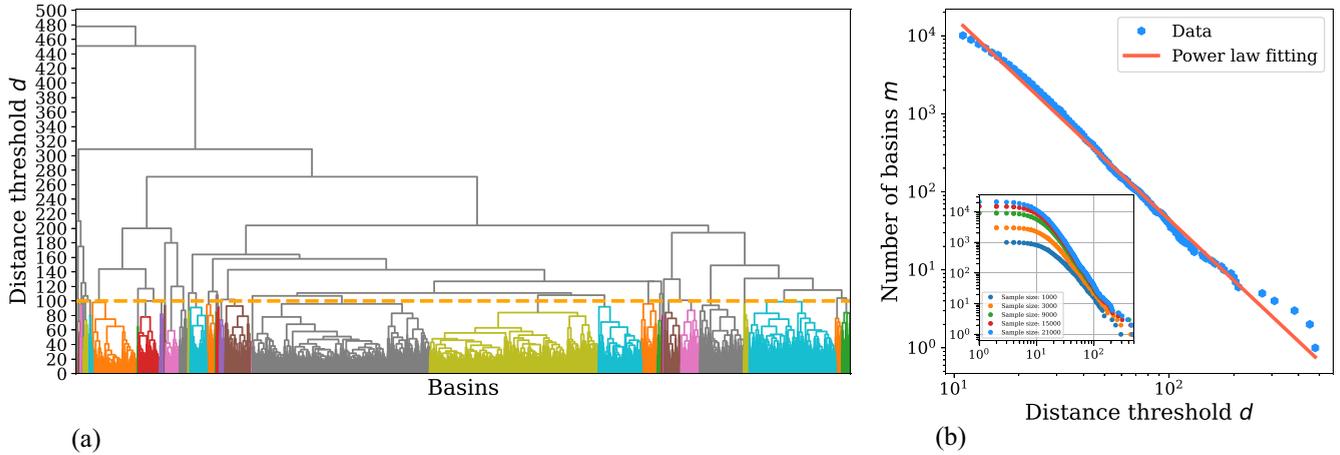

FIG. 1. (a) A dendrogram illustrating the hierarchical clustering of basins, based on a set of 21 000 low-energy states within a small energy window for a system with $N = 958$. Orange dashed line indicates basins counting at $d = 100$. (b) Relationship between the distance threshold $d$ and the number of basins $m$, fitted with a power law (red line). The inset shows convergence of the observed number of basins as the number of low-energy states increases.

How low can the effective temperature [75], $1/\beta_{\text{eff}}$, be? Empirically, we found that

$$\beta_{\text{eff}} \approx 560 \left( \frac{\tau}{20\,\mu s} \right)^{0.16}, \quad (4)$$

in the available range $2\,\mu s < \tau < 2$ ms, where $\tau$ is the annealing time per cycle. Note that this entire range is far from the adiabatic regime, even for our smallest systems with $N$ around 500. The annealing is always performed in the nonadiabatic regime [76]. Yet, the cyclic annealing is capable of "cooling" the system down to a very low effective temperature of $10^{-3}$. We expect that the temperature, decreasing with the increasing annealing time, saturates at a sufficiently large $\tau$ (though we could not confirm it experimentally due to imposed limits). If there are no fundamental limitations on how long such saturation time can be, it seems plausible that $\beta_{\text{eff}}$ can be further increased with an improved hardware and annealing protocols [77]. We thus conjecture that there is no fundamental limit on $\beta$. The situation is reminiscent of the third law of thermodynamics, which precludes reaching zero temperature but does not place limitations on how low the temperature can be.

It is worth noting that exact numerical algorithms for 3D EA model, such as branch-and-cut [1], have been reported [57] to reach the efficiency of $\beta \approx 10^2$ for typical instances. Analog devices, such as D-Wave, can apparently increase it by at least another order of magnitude (admittedly, in a probabilistic rather than in the exact sense). It is likely that the effective temperature can be reduced even further.

The rest of the paper is organized as follows: In Sec. II, we present our results for the number of low-energy states and basins in 3D EA spin glasses. Section III is devoted to the digital "cooling" postprocessing algorithm. In Sec. IV, we describe our results for a set of large size spin glasses. Section V provides a short summary and discussion of the key ingredients of our conclusions.

## II. BASINS AND COUNTING FORMULA

We implemented 3D spin glasses with the Hamiltonian given by Eq. (1) on D-Wave's Advantage 4.1 quantum processor, where $J_{i<j}$'s are independently drawn from the uniform distribution over the interval $[-1, 1]$. The Pegasus architecture [78], denoted by $P_M$, was used for various system sizes. This architecture is a 3D cubic lattice $(M − 1) \times (M − 1) \times 12$, with two spins per unit cell. The data presented in this section were generated using standard forward annealing protocol.

In spin glasses, a basin represents a set of states with similar energies and Hamming distances between any two of them below a certain threshold, denoted by $d$. To visualize and organize the basins, we used dendrograms [see Fig. 1(a)]. The horizontal axes here label 21 000 low-energy states collected for a system size $N = 958$ within a specific narrow energy window. We employed the complete linkage method from SciPy library [79] to generate basins. Each horizontal line in the dendrogram represents a distance threshold $d$; all states connected to it form a basin with the threshold $d$. Figure 1(b) shows the number of basins $m(d)$ at a given distance $d$, i.e., a number of vertical lines intersected by a horizontal line [e.g., the orange dashed line in Fig. 1(a)] at a height $d$. One observes that $m(d)$ follows a power law scaling:

$$m(d) = C\left(\frac{N}{2d}\right)^\alpha, \quad (5)$$

where the fitting parameters are $\alpha \approx 2.6$ and $C \approx 0.76$. Below we show that $C$ is very sensitive to the specific energy window, while $\alpha$ is practically energy independent. One may worry that Eq. (5) is a property of a number of collected states. The inset in Fig. 1(b) shows that adding more states within the same energy window adds extra basins at smallest $d$'s, while larger $d$'s quickly saturate to the relation (5) [see Fig. 2(a)] [80].

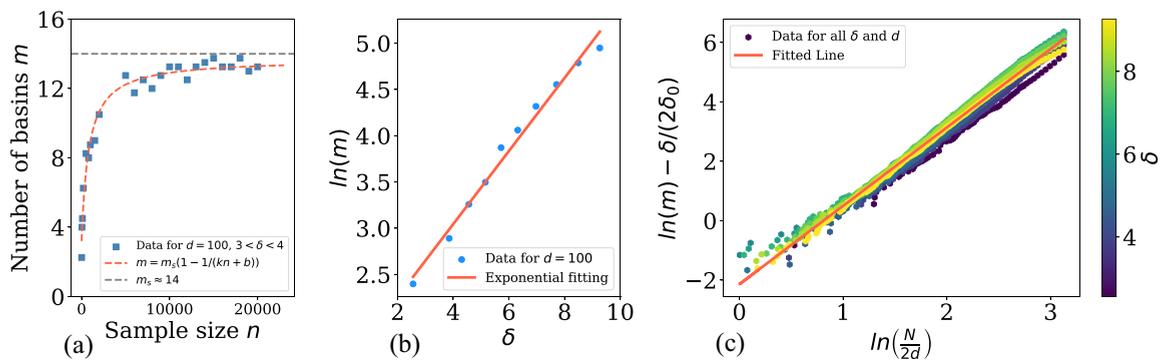

FIG. 2. (a) Number of observed basins (with threshold $d = 100$) as a function of sample size $n$. States are sampled from a fixed energy window $3 < \delta < 4$. Each data point is averaged over 40 trials with same sample size. The dashed red line is a fit to $m = m_s[1 - 1/(kn + b)]$, where $m_s$ is given by Eq. (6), $k = 1.3$, and $b = 1.6$. It shows a number of observed basins saturate with large sample size. (b) Exponential relationship between $\ln(m)$ and $\delta = E - E_0$. The red line represents the exponential fit. (c) Data collapse on Eq. (6) for all values of $\delta$ and $d$ when plotting $\ln(m) - \delta/(2\delta_0)$ vs $\ln(N/2d)$. Color of data points encodes values of $\delta$, as shown on the right.

To investigate the energy dependence of Eq. (5), we generated ten groups of states that ranged from the lowest to a relatively high energy. Each group has 21 000 states within a specific small energy window. We then construct the dendrograms and repeat the counting procedure, described above. The results are presented in Figs. 2(b) and (2c) and are summarized with the best fit:

$$m(\delta, d) = C_0 \exp\left\{\frac{\delta}{2\delta_0}\right\}\left(\frac{N}{2d}\right)^\alpha. \quad (6)$$

Here, $\delta = E - E_0$, where $E$ is the center of the energy window and $E_0$ is the ground-state energy (a way we determine $E_0$ is discussed below), $\delta_0 = |E_0|/N \approx 1.6$, and $C_0 \approx 0.08$.

Focusing on the lowest energies, $\delta \sim O(1)$, one may ask to how many distant, $d \sim O(N)$, basins do such low-energy states belong? According to Eq. (6), the typical answer is *one*. There is typically a *single* basin, which contains both the ground state and all (or most) of low-energy excited states within the energy window $E_0 < E < E_0 + 1$. On the other hand, both the total number of states and the number of distant basins, they belong to, grow exponentially once $E - E_0 \gg \delta_0$. The proliferation of the number of distant basins for different energy levels above the the ground state is illustrated schematically in Fig. 3.

### III. DIGITAL COOLING TECHNIQUE

As discussed in the previous section, the number of distant basins grows exponentially with energy. Yet, if $\delta$ is not too large, there is still a finite number of them. Suppose this is the case and one can generate sufficiently many low-energy states to cover *all* the distant basins and, moreover, have sufficiently many states falling within each big basin. It appears that this is sufficient to recover the true ground state.

The core idea is based on the fact that each basin in the spin glass landscape has an ancestor—a state with lowest energy, from which all other states are generated. All the excited states within such a basin are results of flipping a number of relatively limited porous clusters of spins with a small surface energy. The idea thus is to identify such loosely connected clusters within each basin. By selectively flipping individual clusters, one can proceed lowering the energy until the bottom of the basin is reached. The procedure, outlined below, directs the process toward the single deepest basin and thus converges to the true ground state.

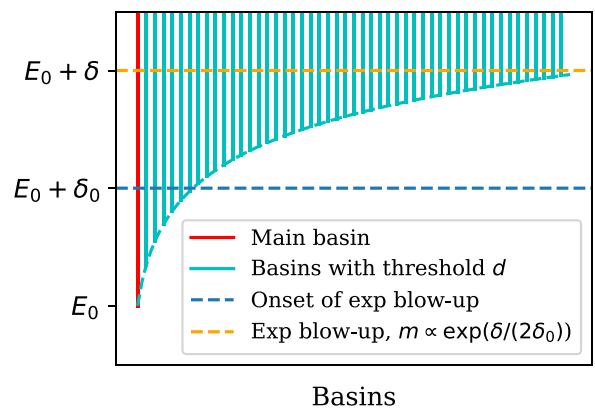

FIG. 3. Schematic illustration for proliferation of distant basins. The main basin (in red) corresponds to the region near the exact ground state, which has energy $E_0$. As the energy increases, basins (in cyan) proliferate. Dashed lines illustrate the number of basins at each energy level $E = E_0 + \delta$.

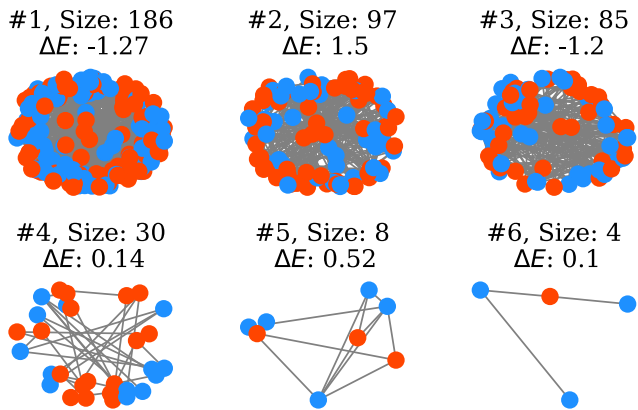

FIG. 4. Examples of connected clusters with small surface energy between two low-energy states with a total Hamming distance of about 450. Six largest clusters are shown here. Blue nodes denote up-spin, while red nodes denote down-spin.

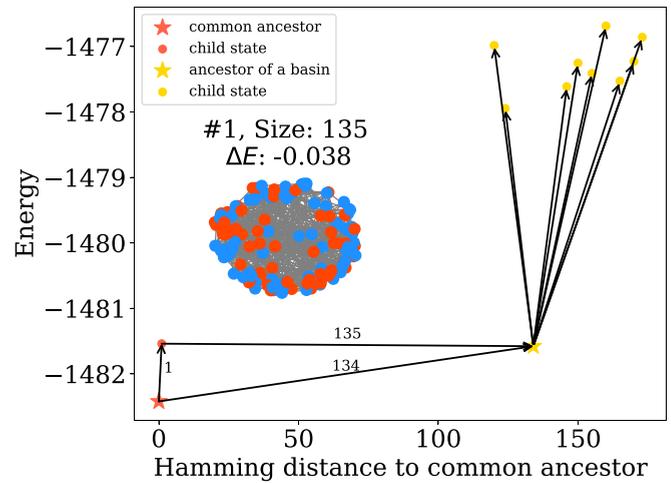

FIG. 5. Illustration of the exact ground state acting as a common ancestor (red star). A distant ancestor (yellow star) is generated from the common ancestor by flipping a 134-size cluster or equivalently by flipping a single spin first to a child state (red dot) and then flipping a 135-size nearly-zero-energy cluster (shown in inset). States in a basin (yellow dots) are generated from the ancestor (yellow star) through some small-size clusters' flipping.

To identify such clusters, we calculate a Hamming distance $d$ between a pair of low-energy states. This implies that $d$ spins should be flipped to go from one of those states to the other one. One now focuses on those $d < N/2$ spins, which are different between the two states, and looks whether they are connected on the lattice through nonzero $J_{ij}$ couplings. This splits the $d$ spins into a several connected clusters, such that all spins belonging to different clusters do not have any nonzero couplings between them. Examples of such clusters with their sizes and surface (i.e., flip) energies are illustrated in Fig. 4; a statistical analysis of these characteristics may be found in Ref. [74]. Notice that, despite large sizes of some of these clusters, the energy change of flipping the entire cluster is rather small, $O(1)$. Such clusters may be digitally flipped independently from each other to generate new low-energy states.

The digital cooling technique works as follows: consider a set of low-energy states produced by the annealer. Let us denote them as $a, b, c, d, \ldots$ Pick one state, say $a$, and compare it pairwise with states $b, c, d, \ldots$, as discussed above. This identifies a library of clusters, which may be flipped within state $a$ resulting in small energy changes. Choose now one such cluster, which promises the largest energy decrease, and flip it, producing a new state $a'$. By construction, the energy of $a'$ is less than that of $a$. One can now repeat this procedure comparing $a'$ with $b, c, d, \ldots$, identifying a cluster with largest energy decrease and producing an even lower energy state $a''$, by flipping this cluster. One proceeds this way until *all* the clusters in the state $a'' \cdots'$ can only increase its energy. This leads to a new state $a_1[a; b, c, d, \ldots]$, which we call a parent of $a$. Indeed, the state $a$ is obtained from its lower energy parent $a_1$ by flipping a number of clusters. One then turns to the state $b$ and compares it pairwise with $a, c, d, \ldots$ The result is its parent state $b_1[b; a, c, d, \ldots]$, etc. Proceeding this way, one obtains a new set of states $a_1, b_1, c_1, d_1, \ldots$ with energies lower than the original states. Moreover, it appears that states in such parental set are occasionally identical, e.g., $b_1 = c_1$. This implies that distinct states, $b \neq c$, have the common ancestor, $b_1$. Thus, the ancestral set $a_1, b_1, d_1, \ldots$ is smaller and deeper in energy than the original one. One then produces a next ancestral generation $a_2, b_2, \ldots$, which is again smaller and deeper. One proceeds this way until a *single* common deepest ancestor, $a_k$, remains after $k$ generations. In practice, the process usually converges in 2–3 generations. Somewhat similar algorithms were employed in Refs. [40–42].

The procedure guarantees that each deeper parental generation has energies smaller than their kids. According to Eq. (6) and Fig. 3, this implies fewer basins available for such lower-energy set of states. As a result, Fig. 3 works like a sink, which directs the algorithm toward the unique ground state. Given a sufficiently large initial set, the common ancestor state must be the ground state. To accelerate the processing, we divide states into groups of, say, 20 and find their common ancestor. Often, it is the same for *all* the groups, giving a highly probable ground state. If not, one proceeds to looking for the next generation of ancestry, until all groups of states collapse to the exactly same common ancestor.

Figure 5 illustrates an example of such genealogy tree, which may be reconstructed with the cluster flipping procedure. The larger clusters play a special role by moving the common ancestry search from one distant basin to another. The exact ground state is the common ancestor of all states. An offspring is generated from it by flipping a large cluster with a small energy. In Fig. 5, a higher-energy state is generated by flipping a specific spin in the common ancestor. Then, the offspring is generated by flipping a cluster of size 135 and energy $-0.038$. This state is an ancestor of a local basin (denoted by yellow), from which a number of daughter states are generated through small clusters flips. In practice, the ancestry search runs in the opposite direction: from the top to the bottom.

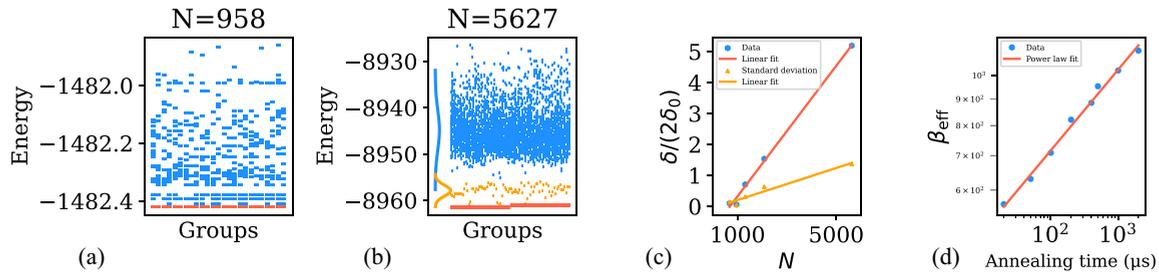

FIG. 6. (a) For $N = 958$, 500 low-energy states (blue bars) are divided into 25 groups. Common ancestor states, generated through the digital cooling, are shown with red bars. They are the same for all 25 groups, with the energy $-1482.421$. (b) For $N = 5627$, 4000 low-energy states (blue bars) are divided into 80 groups. The orange lines represent common ancestors in each group. Curves on the left show the energy distributions before and after digital cooling. The orange states were divided into two groups, yielding two very close common ancestors (red lines). (c) Average energy $\delta/(2\delta_0)$ (blue dots) and standard deviation $\sigma$ (orange dots) as functions of the system size $N$ for a fixed cyclic quantum annealing protocol. The linear fit (red) $\delta/(2\delta_0) = 0.00105N - 0.733$. (d) The inverse average residual energy per spin, $\beta_{\text{eff}}$, as a function of the annealing time per annealing cycle. All other parameters are fixed. A power law relation (4) is found by the linear fit of the log-log plot.

## IV. FINDING GROUND STATES OF LARGE SPIN GLASSES

To illustrate the procedure, we discuss its details for 3D spin glass systems of sizes $N = 958$ and $N = 5627$. It starts from running the cyclic annealing [73] with random initial states to generate a large ensemble of low-energy states. Details of the cyclic annealing protocol and its practical implementation can be found in Ref. [74] and the Appendix. It mimics a cooling engine cycle. First, it drives the spin glass into a phase with a unique *gaped* ground state by selectively biasing the energy toward one specific *reference* state. It then increases the transfer field to reach a paramagnetic (still gaped) ground state. Finally, it completes the cycle by simultaneously decreasing both the bias and the transverse field. This latter step brings the system back into the gapless spin glass phase, arguably through the second-order transition of the many-body localization type. The cycle is exothermic and thus a measurement, taken upon its completion, may result in a lower-energy state than the initial reference state. If indeed, such a state is taken as the new reference state and the cycle is repeated until it stops producing lower-energy states. The process then starts over from a random initial reference state. After collecting sufficiently many of such cycle-termination states, we randomly divide them into smaller groups and run the digital cooling (ancestry search) algorithm within each group.

In the case of $N = 958$, we repeatedly found *exactly* the same common ancestor state in each of the independently generated groups. Figure 6(a) shows the data for a random typical realization of the glass; 500 low-energy states (denoted by blue bars) were generated via cycle annealing with random initial states. They are then divided into 25 groups, each containing 20 states. Each group corresponds to a column in the figure, and the common ancestor state from each group (after digital cooling) is marked with a red bar. All 25 groups produced *exactly* the same state, with the energy $-1482.421$. Evidently, the probability this state is *not* the true ground state approaches zero exponentially with the number of groups, yielding the same ancestor. This scenario was consistently reproduced in several random realizations of $N = 958$ EA spin glasses.

In case of $N = 5627$, we generated 4000 low-energy states through the cyclic annealing with random initial states. Those states were randomly split into 80 groups of 50 states each, blue bars in Fig. 6(b). After applying the digital cooling, a common ancestor for each group was generated. In this case, most of these common ancestors do not coincide, orange lines in Fig. 6(b). Their energies form a distribution centered at about $-8958$ with the standard deviation $\sigma_1 = 1.16$ [to compare the initial (blue) states were centered at $-8944$ with the standard deviation $\sigma_0 = 4.39$].

We then divide a set of 80 common ancestor states into two second-generation groups. After applying digital cooling to each of these two groups, we found them yielding two very close (yet still different by a single cluster of 67 spins) common ancestors with extremely close energies $-8961.40$ and $-8961.11$, red lines in Fig. 6(b). We believe the first one is the true ground state, though the confidence level of this assertion is much less than for $N = 958$.

To investigate dependence of the computational complexity on the system size, we kept the cyclic quantum annealing settings fixed (same annealing time, protocol, and number of cycles) and tested system sizes $N = 678, 958, 1312, 2084,$ and $5627$. The cyclic annealing with random initial conditions produces a narrow Gaussian-like distribution of energies. Figure 6(c) shows that both the average residual energy $\delta/2\delta_0$ and the standard deviation $\sigma$ scale approximately linearly with the system size $N$, with slopes of $10^{-3}$, Eq. (3), and $2.6 \times 10^{-4}$, respectively. To know the absolute scale of the excess energy, one needs to know the ground-state energy for each system size. The digital cooling algorithm was run for each system size. Except for the largest case of $N = 5627$, it rapidly converges to an exactly same common ancestor. Its energy was taken as $E_0$.

The linear relationship (3) may be interpreted as a finite density per volume of disconnected clusters with a smooth distribution of $O(1)$ energies. The number of excited (flipped) clusters per unit volume is dictated by an inherent noise of the nonadiabatic annealing process. The magnitude of such noise, and thus $\beta_{\text{eff}}$ in Eq. (3), may be controlled by changing the annealing rate. Figure 6(d) shows $\beta_{\text{eff}}$ as a function of the annealing time per cycle, $\tau$, with the fit given by Eq. (4).

## V. DISCUSSION AND CONCLUSION

For completeness, let us discuss here an algorithm producing the subexponential complexity, $\sim 2^{4N^{2/3}}$ [81]. Consider a cube of size $L = N^{1/3}$ and let us assume that its ground state can be found with a complexity $2^{f(L)}$. Let us now slice this cube in three orthogonal directions onto eight cubes of size $L/2$. Given a fixed configuration of spins sitting on the three cutting planes, the eight cubes are completely *uncoupled* thanks for the nearest-neighbor interactions of the EA model. Therefore, for every of $2^{3L^2}$ configurations of the spins on the cutting planes, one needs to determine a ground state for each of the eight cubes of size $L/2$ independently. To this end, we will cut each of $L/2$ cubes into eight cubes of size $L/4$, etc. Complexity of this procedure is thus $2^{3L^2} \times 8 \times 2^{f(L/2)}$, which leads to the recursion relation

$$f(L) = 3L^2 + 3 + f(L/2). \tag{7}$$

Easy to see that it is solved with $f(L) = 4L^2 + 3\log_2 L$, resulting in the complexity $N \times 2^{4N^{2/3}}$. It is possible that with a smarter bookkeeping, the factor of 4 in the exponent can be somewhat reduced. However, it cannot be less than 1, according to the lower bound on the complexity, proven in Ref. [55], under the assumption of ETH [56] validity. Our point here is to demonstrate existence of a subexponential, $N^{2/3}$, algorithm—not to prove the lower bound.

Based on the data obtained with the D-Wave Advantage annealer, we observed an exponential scaling, $2^{N/\beta}$, which is, of course, inferior in the limit $N \to \infty$. Yet, we found $\beta \approx 10^3$, and argued that it is feasible to increase it even further. This implies that for $N < \beta^3$, the annealer with its exponential scaling is more efficient than the subexponential algorithm. (For $N > \beta^3$, the subexponential method requires a computational time in excess of $2^{\beta^2}$, which is not viable even for $\beta = 10$.)

The venue to reach a very large $\beta$ is not universal across optimization tasks, or even spin glass models. There are reasons to believe that it is restricted to short-range, spatially local models (so is the subexponential algorithm, as well). Indeed, it is based on the idea of independent, relatively small, disconnected from each other excitation clusters. This, of course, does not work for the all-to-all Sherrington-Kirkpatrick (SK) model. Thus, no large $\beta$ was ever reported for SK model. The largest, we are aware of, is $\beta = 1/0.226 = 4.42$, recently proved for a quantum algorithm [58]. While it is possible that it will be somewhat increased, one may expect that there is a fundamental limit on it.

Finally, we address an issue of whether the quantum nature of the annealer is important for these conclusions. In our opinion, it is *not*. Both classical and quantum annealing can, in principle, reach a low average residual energy per spin, $1/\beta_{\text{eff}}$, Eq. (3). One may argue that low effective temperature requires reduction of the internal noise and thus reduction of the physical temperature, where the quantumness inevitably shows up. It is thus plausible that quantum devices are capable of reaching larger $\beta$'s than classical ones. On the other hand, one may think of "quantum-inspired" algorithms, e.g., where coupled spins evolve in transverse fields according to the classical Landau-Lifshitz-Gilbert equation [45]. To the best of our understanding, it is not clear whether such fully classical algorithm is inherently inferior to the nonadiabatic quantum annealing.

Several open questions remain. One key question is how to improve $\beta_{\text{eff}}$ in practice, whether through advancements in cyclic quantum annealing [73,74], improvements in quantum devices [53], or the implementation of exotic quantum drivers [82]. Another open question is whether particularly hard instances exist beyond random realizations for large spin glasses.

## ACKNOWLEDGMENTS

We are grateful to Mohammad Amin, Dmitry Bagrets, Andrew Berkley, Tim Bode, Emile Hoskinson, David Huse, Liang Jiang, Andrew King, Giorgio Parisi, Jack Raymond, Boris Shklovskii, and Michael Winer for insightful discussions. This work was supported by the NSF Grant No. DMR-2338819.

## DATA AVAILABILITY

Some of the data that support the findings of this article are openly available [83]. Additional data analyzed are not publicly available. These data are available from the authors upon reasonable request.

## APPENDIX: CYCLIC QUANTUM ANNEALING

Here, we briefly review the cyclic quantum annealing protocol. Traditional forward quantum annealing [84] follows an open path connecting a driver Hamiltonian $H_q$, with an easily prepared ground state, to a problem Hamiltonian $H_p$. This is illustrated by the blue path in Fig. 7. Cyclic annealing [73,74], on the contrary, traverses a closed path in the parameter space that passes through the problem Hamiltonian, shown as the orange path in Fig. 7. Such a cyclic process enables progressive energy reduction, cycle by cycle, until a sufficiently low-energy state is reached. In practice, it can save up to 85% of the annealing time compared to forward annealing to reach states with a similar energy.

The algorithm, employed here, is based on the following Hamiltonian:

$$H(s) = H_p + B_z(s)H_{\text{ref}} + B_x(s)H_q, \tag{A1}$$

where $s$ parameterizes the cycle and the problem Hamiltonian is

$$H_p = \sum_{ij}^N J_{ij}\, \sigma_i^z \sigma_j^z, \tag{A2}$$

with coupling parameters $J_{ij}$ encoding an optimization problem at hand.

The protocol includes three steps, each serving a specific function: *Step 1—biasing toward a reference state:* The device is initiated into a certain bit-string *reference* state, $\sigma_i^z = s_i$, typically chosen at random. One then increases $B_z(s)$ coupled to the classical reference Hamiltonian of the form

$$H_{\text{ref}} = -\sum_i^N s_i \sigma_i^z. \tag{A3}$$

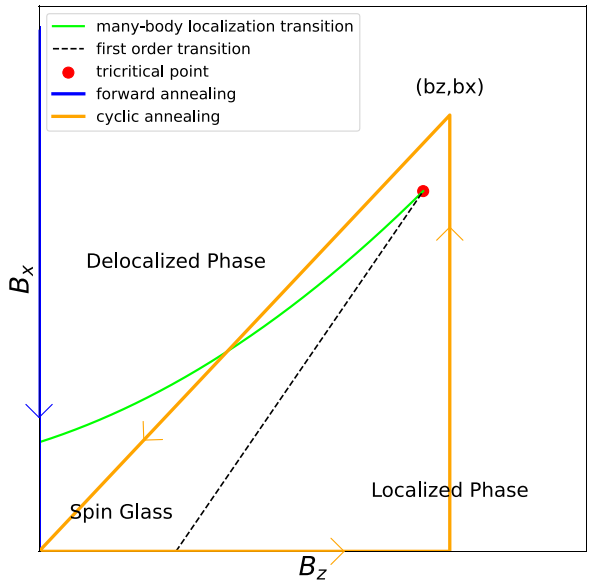

FIG. 7. Schematic phase diagram. Schematic representation of the phase diagram for the system defined in Eq. (A1). The green line marks the many-body localization transition. The black dashed line denotes the first-order transition between gapless spin glass phase and a phase with a gaped ground state, many-body localized around the reference state. The path taken by the cyclic annealing protocol is illustrated by the orange triangle, contrasting with the blue path followed in standard forward annealing.

This biases the system toward the reference state $\{s_i\}$, passing through the first-order transition. At this point, the reference state becomes the true ground state of the combined Hamiltonian $H_p + B_z(s)H_{\text{ref}}$. *Step 2—turn on the quantum driver Hamiltonian:*

$$H_q = -\sum_i^N \sigma_i^x. \quad (A4)$$

This term does not commute with $H_p$ or $H_{\text{ref}}$ and introduces quantum fluctuations, enabling delocalization and superpositions of bit-string states. For sufficiently large $B_x$, it drives the system through the many-body localization transition. This step is executed by increasing $B_x(s)$ from 0 to $b_x$. Due to the bias of step 2, the system mixes primarily with states that have lower $H_p$ energy than the reference state. In D-Wave's device, a reasonable choice is $b_z = 0.03$, $b_x = 8.04$. *Step 3—complete the cycle and perform a measurement:* In the final step, $H_{\text{ref}}$ and $H_q$ are simultaneously turned off, returning the system to the problem Hamiltonian and completing the cycle. A measurement is then performed, collapsing the state into a classical bit string. While the final state is generally not the ground state—due to Landau-Zener transitions during the turn-off process—the bias from $H_{\text{ref}}$ increases the probability of obtaining a state with lower energy than the initial reference. If this happens, we use the so-found lower-energy state as the new reference state and repeat the cycle. Otherwise, we reinitialize into the previous reference state and again repeat the cycle. The cycle is run until a certain number of successive attempts (say 5) fail to produce a lower-energy state. At this point, the state is recorded and one starts over from a new randomly chosen reference state. More details can be found in Ref. [74].